\documentclass[aps,prb,superscriptaddress,twocolumn,showpacs,floatsfix]{revtex4-2}
\usepackage{float}
\usepackage[french,english]{babel}
\usepackage[T1]{fontenc}
\usepackage[utf8]{inputenc}
\DeclareUnicodeCharacter{2212}{-}

\usepackage{graphics,graphicx,caption,subcaption}
\usepackage[dvipsnames,usenames]{color}
\usepackage{rotating}

\usepackage{amssymb,amsmath}
\usepackage{wasysym}
\usepackage{xfrac,nicefrac}
\usepackage{gensymb}
\usepackage{bm}
\usepackage{pst-node}
\usepackage{relsize}

\usepackage{units}

\usepackage[active]{srcltx}
\usepackage[sort&compress]{natbib}

\usepackage{hyperref}
\hypersetup{
  colorlinks=true,
  citecolor=blue,
  linkcolor=blue,
  urlcolor=blue,
}

\usepackage{array}
\usepackage{tabularx}
\usepackage{multirow} 
\usepackage{dcolumn}

\usepackage[color=orange!20]{todonotes} 
\usepackage[normalem]{ulem}
\usepackage{soul}

\newcolumntype{d}[1]{D{.}{.}{#1}}

\newlength{\dbarheight}

\newlength{\ddbarheight}

\newcommand{\ybmo}{YbMnO$_3$}
\newcommand{\lumo}{LuMnO$_3$}

\newcommand{\scmo}{ScMnO$_3$}
\newcommand{\dymo}{DyMnO$_3$}
\newcommand{\rmo}{RMnO$_3$}
\newcommand{\bcs}{Ba$_2$CoS$_3$}
\definecolor{okgreen}{rgb}{0.10,0.37,0.13}

\definecolor{verifyblue}{HTML}{1F4E79}
\definecolor{tealc}{HTML}{0E7C7B}
\definecolor{analogypurple}{HTML}{5E2D7B}
\definecolor{cautionorange}{HTML}{ffa204}
\definecolor{alertred}{HTML}{B22222}
\definecolor{spincol}{HTML}{0000CD}   
\usetikzlibrary{arrows.meta,shadows,patterns}
\definecolor{capgray}{gray}{0.38}
\DeclareCaptionFont{capsize}{\fontsize{6}{7.2}\selectfont}
\begin{document}
\title{%
Small-data physics-informed ML model for magnetic exchange based on 
      chemical composition and structure
}

\author{Brice Kengni Zanguim}
\affiliation{Institut N\'eel, CNRS UPR 2940,  Grenoble, France} 

\author{Thibaut Véry}
\affiliation{Institut du Développement et des Ressources en Informatique Scientifique, CNRS UAR851, Orsay, France}

\author{Marie-Bernadette Lepetit}
\affiliation{Institut N\'eel, CNRS UPR 2940,  Grenoble, France} 
\affiliation{Institut Laue Langevin, Grenoble, France}


\begin{abstract}
  In this paper we propose an ensemble model based on a swarm of dense neural networks,
  able to predict Heisenberg exchange interactions within less than 1\%
  error. The model  uses the inputs of the ab-initio Schrödinger equation,
  i.e.  the compound structure and chemical composition. 
  In this first work, we trained the model on a few compounds of the hexagonal manganite family and
  were able to infer, with similar accuracy, the $J$ values for other compounds
  of the same family at randomly chosen geometries, thus showing that we
  accurately learned the effect of the chemical composition and structure on
  effective magnetic integrals $J$.  \pacs{}
\end{abstract}

\maketitle

\section{Introduction}
\label{sec:intro}


Strongly correlated quantum materials, such as transition-metal oxides and
$f$-electron systems, display remarkable electronic and magnetic properties
(superconductivity, magnetism, topological order) that arise from the complex
interplay of electron--electron interactions, spin--orbit coupling and lattice
effects.

The consequence of these interplay is that these properties cannot be
described by conventional single-particle theories.  The standard approach is
to build an effective many-body low-energy Hamiltonian (for instance an
Heisenberg~\cite{Heisenberg1928} or Hubbard~\cite{Hubbard1963} model) that
captures the relevant physics. These simplified models give access to ground
states, phase diagrams and low-energy excitations, either through standard
condensed-matter methods (exact diagonalization, Dynamical Mean-Field Theory
(DMFT)~\cite{DMFT_Pavarini2014}, Projected Entangled Pair States
(PEPS)~\cite{PEPS_Cirac2021}, Monte Carlo, etc.) or, more
recently, deep-learning approaches~\cite{Carrasquilla2017,Carleo2017}.\\

The associated effective interactions, such as the $U$ and $t$ of a Hubbard's
Hamiltonian or the magnetic exchange integrals, $J$, of the Heisenberg one, are
intrinsically dependent on both the atomic structure and the chemical
composition of the compounds. As a consequence they provide insight to the
composition-structure-properties relationship. Accurate and fast methods to obtain these
effective interactions for given compounds and geometries are thus of
importance, not only for the understanding of the physical properties and
their grounding in the structure and chemistry of the materials, but also for
computational material design.\\

Even if Density Functional Theory (DFT) methods are often used to evaluate the
effective interactions, they lack accuracy and reliability due to their
intrinsic difficulty to describe strong electronic correlation. To reach those
goals one thus has to rely on post-Hartree-Fock multi-reference (MR)
wave-function methods. The Difference Dedicated Configuration
Interactions~\cite{DDCI1992,DDCI1997} (DDCI), and the Selected Active Space +
Single excitations~\cite{p:SASS,p:relaxse} (SASS) methods have been
specifically designed for this purpose and have proven their ability to reach
experimental accuracy~\cite{p:YMnO3_3,J_Cu_CompMth_2023}.\\

Ab-initio methods, whether of MR-wave-function type or DFT, are however
computationally demanding and rapidly become prohibitive, for instance when
associated with material design algorithms. This is a typical case where
artificial intelligence models can provide new avenues. Few papers are however
tackling this specific problem, and even less try to directly evaluate the
effective magnetic exchanges from the chemical composition and structure only.
One should nevertheless cite the work of J.~Bahlke {\it et
  al}~\cite{Bahlke_J_2020} where the authors explored the learnability of
magnetic exchange interactions in di-copper complexes by Gaussian process
regression. They found that the use of internal coordinates such as distances
and angles out-performed more sophisticated descriptors such as the Smooth
Overlap of Atomic Positions (SOAP)~\cite{Bartok2013}, the Many-Body Tensor
Representation (MBTR)~\cite{MBTR_Rupp2022_1,MBTR_Rupp2022_1} or the Many-Body
Interaction ($\rm F_{MB}$\!)  Descriptors~\cite{FMB_Muller2018}.  They
however worked on molecular complexes, i.e. on small finite-size systems where
the impact of distant atoms does not have to be accounted for.\\

In this paper we propose an ensemble model, built upon several dense neural
networks (DNN), to predict the exchange integrals in strongly correlated
quantum materials from structure and chemical composition. In this first work
the model was trained on the hexagonal manganites \rmo\ (R being a
rare-earth). Not only the model can reach accuracy below 1\% on the $J$
values, but it is able to
predict the magnetic exchange on unknown compounds of the same family.\\

The next section will be devoted to the model description,
section~\ref{sec:data} to the database building and section~\ref{sec:dataE} to
the data engineering. Section~\ref{sec:res} will describe the model
hyper-parameters and results, while the last section will discuss and conclude
this study.

\section{The model }
\label{sec:modele}

Most machine learning models are based on large datasets. Indeed, with such
datasets, regression processes can learn information deeply buried in the
provided descriptors, i.e. depending in a very indirect way on the provided
data. In the present case, not only the data are not already available, but
the datasets construction is quite human and computationally demanding, as
already pointed out. In the small data regime the main danger is over-fitting
or memorization of the data (and the noise in the latter). In order to avoid
such problems one thus needs to use only few parameters in the machine
learning (ML) models, with the corollary that small ML models encounter more
difficulties to learn information too deeply embedded in the data.  In such
cases the learning success depends mostly on the closeness of the final
descriptors to the ML-model desired output. In other words, the physical
knowledge should be embedded in the data engineering.

In our case the effective exchange interactions can be deduced from the
knowledge of the local ground and excited states of the system, i.e. from the
eigenvalues of the local Born-Oppenheimer ab-initio electronic
Hamiltonian~\cite{p:revue,DDCI1992,p:SASS}. All the required information is
thus embedded in the parameters of this Hamiltonian, that is in the location
and chemical nature of the atoms involved in the seek interactions and the
embedding potential of this crystal
fragment~\cite{p:env,p:env2,EmbedAIMP_Seijo,Scepic_2022}. While those raw
descriptors could be used in a large model, trained on large datasets, the
seek magnetic exchange interactions are embedded to far away in them. We thus
need to engineer the data so that the inner descriptors passed to the neural
network are close enough to the physical or chemical dependency of the
magnetic exchanges.  For this purpose we have a few guides such as the
Kanamory-Goodenough rules~\cite{Goodenough1963} and the second-order
quasi-degenerate perturbation theory~\cite{J_QDPT} from which they can be
derived.

Most ML-models for material properties are based on Graph Neural Networks
(GNN)~\cite{Xie2018,Schutt2017,Batzner2022,Batatia2022,Gong2023} as they
provide a natural representation of the crystal. Indeed, the GNN structure can
be directly mapped onto the parameters of the underlying Schr\"odinger
equation, with the nodes representing the atoms and the adjacency matrix
representing the inter-atomic distances, coulomb potential or inter-atomic
properties. While this description is perfectly adapted to the learning of the
global properties of the system, it is less suited when one is interested in
local properties such as magnetic exchange
interactions~\cite{Localite_J_1,Localite_J_2,p:revue}. Indeed a GNN model is
best suited when all atoms have an equal importance, while the present problem
can be structured in a quantum fragment, including the atoms involved in the
interaction, and its embedding, a construction not natural to GNN
architectures, that therefore need to be learned by the model.  This learning
would requires large datasets that we cannot afford, while small Dense Neural
Networks (DNN) learning on engineered internal descriptors do not require
large datasets, as done by Behler and Parrinello~\cite{Behler2007}.

In order to determine the best model we did the training with a set of
different hyper-parameters such as~: the number of hidden layers, the latent
space dimension, the hyper-parameters describing the embedding, the activation
functions at each DNN layer, the loss function. We of course 
trained our different models with several random seeds.  We then applied an
iterative clustering method in order to select the best models over a set of
quality functions on the training and testing subsets. Finally we designed an
ensemble model on the best models obtained in the last clustering iteration.

We tested our final model, not only on the train/test data subsets, but also
on an \rmo\ compound that was excluded from the learning process from
start. This last steps allow us to test the transferability, i.e. ability to
extrapolate, of the model within the chosen material family.

\begin{figure*}[tp]
\centering
\resizebox{\textwidth}{!}{%
\begin{tikzpicture}[
   >={Stealth[length=3.4mm,width=2.8mm]},
   font=\small,
   stage/.style={draw, line width=1pt, rounded corners=7pt, align=center,
                 text width=3.0cm, minimum height=1.75cm, inner sep=5pt,
                 drop shadow={opacity=0.18, shadow xshift=1.3pt, shadow yshift=-1.7pt}},
   badge/.style={circle, draw=white, line width=0.7pt, fill=#1, text=white,
                 font=\bfseries, inner sep=0pt, minimum size=7.25mm},
   arr/.style={->, line width=1.6pt, draw=black!45}]

  \node[stage, draw=verifyblue,    top color=verifyblue!4,    bottom color=verifyblue!20]    (d)  at (0,0)  {\textbf{Data}\\[2pt]\small Atomic data \& $J$};
  \node[stage, draw=tealc,         top color=tealc!5,         bottom color=tealc!24]         (de) at (4,0)  {\textbf{Data engineering}};
  \node[stage, draw=analogypurple, top color=analogypurple!5, bottom color=analogypurple!22] (m)  at (8,0)  {\textbf{Dense networks}\\[2pt]\small $\sim$\,35\,000 trainings};
  \node[stage, draw=cautionorange, top color=cautionorange!5, bottom color=cautionorange!24] (c)  at (12,0) {\textbf{Clustering}\\[2pt] best 138 models};
  \node[stage, draw=alertred,      top color=alertred!4,      bottom color=alertred!18]      (e)  at (16,0) {\textbf{Ensemble model}};
  \node[draw=okgreen, line width=1.1pt, circle, fill=okgreen!16, minimum size=1.35cm] (j) at (19,0) {\large$\widehat{J}$};
  \node[stage, draw=cautionorange, top color=cautionorange!5, bottom color=cautionorange!24] (t)  at (22,0) {\textbf{Transferability test}};

  \node[badge=verifyblue]    at (d.north west)  {1};
  \node[badge=tealc]         at (de.north west) {2};
  \node[badge=analogypurple] at (m.north west)  {3};
  \node[badge=cautionorange] at (c.north west)  {4};
  \node[badge=alertred]      at (e.north west)  {5};

  \foreach \a/\b in {d/de, de/m, m/c, c/e, e/j, j/t} \draw[arr] (\a) -- (\b);
\end{tikzpicture}}
\caption{Model pipeline, from the atomic data to the predicted exchange coupling~$\widehat{J}$, followed by transferability test on unknown compound.}
\label{fig:pipeline}
\end{figure*}
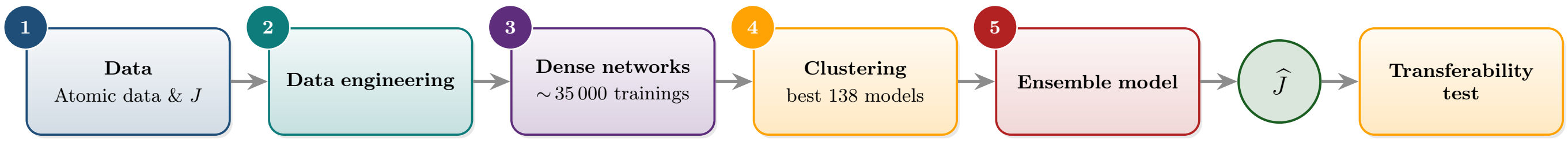

Summarizing the above discussion we came to the workflow pictured in
Fig~\ref{fig:pipeline}.

\section{The database}
\label{sec:data}

In this first work we focus on a specific material family, namely the
hexagonal manganites. We built datasets for \rmo\ compounds with R= Sc, Dy,
Yb, Lu and a completely different compound, namely Ba$_2$CoS$_3$ to prevent
too much specialization of the ML model. We generated 100 different geometries
per $J$ value, randomly distributed around the experimental
ones~\cite{ScMnO3_stuct_xu1995,DyMnO3_stuct_remsen2011,
  YbMnO3_Struct_VanAken2001,LuMnO3_Struct_VanAken2001,Neutron_Scat_Ba2CoS3}. While
the unit cell were preserved, as well as the $\alpha$, $\beta$, $\gamma$
angles, the lattice parameters as well as the fractional atomic positions of
all atoms in the unit cell was randomly modified. As a consequence the space
groups were not preserved and are in general $P1$. In order to insure that the
distorted geometries remain chemically reasonable we restrained the
distortions on the lattice parameter to remain within $4\permil$ of their
pristine values, and the distortions on the atomic positions to remain in a
\unit{0.04}{\AA}-radius sphere around their experimental positions.

In order to enforce the best accuracy for the $J$ database, we used a
specifically designed multi-Reference Configuration Interaction approach,
namely the Selected Active Space + Single excitations~\cite{p:SASS,p:relaxse}
(SASS) method to build the database. Table~\ref{tab:Jexp} summarizes the
computed and experimental $J$ values for the experimental geometries,
confirming the quality of our methodology.

\begin{table}[htbp]
\centering\footnotesize\setlength{\tabcolsep}{7pt}
\begin{tabular}{lcccc}
\hline \hline
  Compound & Exp. & \multicolumn{3}{c}{$J$  (meV)}   \\
           & Struct. & This work &  Exp. &  Ref.   \\
  \hline 
  \multirow{2}{*}{\lumo}&\cite{LuMnO3_Struct_VanAken2001} & \multirow{2}{*}{-3.15 ; -3.83} & -1.54 ; -4.09  & \cite{J_LuMnO3_INS_Bourges_2010} \\
                        &                                 &                                & -3.0 ; -3.2    & \cite{J_Oh2016} \\[2ex]
  \ybmo                 &\cite{YbMnO3_Struct_VanAken2001} & -3.38 ; -3.53                  & -2.65          & \cite{th_XF} \\[2ex]
  \scmo                 &\cite{ScMnO3_stuct_xu1995}       & -4.22 ; -5.26                  & ---            & --- \\[2ex]
  \dymo                 &\cite{DyMnO3_stuct_remsen2011}   & -2.86 ; -2.86                  & ---            & --- \\[2ex]
 Ba$_2$CoS$_3$          &\cite{Neutron_Scat_Ba2CoS3} & -3.48                     & -3.19 $\pm$ 0.17 & \cite{Neutron_Scat_Ba2CoS3} \\
\hline \hline
\end{tabular}
\caption{Exchange integral $J$ obtained for the different compounds used in
  the database, computed at experimental geometries. The experimental $J$
  values from the literature were extracted from inelastic neutron scattering,
  All couplings are antiferromagnetic and of the same order of
  magnitude. }
\label{tab:Jexp}
\end{table}

\section{Data Engineering}
\label{sec:dataE}

The Kanamory-Goodenough rules~\cite{Kanamori1959,Goodenough1963} provide the
qualitative dependence of the effective exchange integrals $J$ with respect to
the distances and angles between the atoms involved in the
interaction.
From the learning point of view it means that these distances and angles must
be computed from the atomic positions to be part of the final internal
descriptors.

These parameters however provide only information on the atoms directly
involved in the interaction and neglect completely the effect of the rest of the
crystal. We however know that the environment provided by the rest of the
crystal is crucial, and that the $J$ values depend on it. To exemplify this
importance one can cite the study on the effect of the counter-ions on the
magnetic integrals in cuprate
$\rm (Ca_x La_{1−x})(Ba_{1.75−x}La_{0.25+x})Cu_3O_y$
superconductors~\cite{p:supra_local}, in which it was shown that exchange
integrals can vary by more than 25\% in amplitude, according to the local
arrangement of the Ca, La and Ba counter-ions. Using internal coordinates, as
for the atoms directly involved in the interaction would not be a very
transferable method, and would require the ML model to learn the importance of
each embedding atoms irrespective to their importance/distance to the
center. We thus described the embedding by the rest of the crystal using a Smooth
Overlap of Atomic Positions method~\cite{Bartok2013} (SOAP). Indeed, this
method allows to encode the atomic density, possibly weighted by any property
descriptor, in a unique set of angular and radial coefficients, acting as
internal descriptors. 
\begin{equation}
  \label{eq:SOAP}
c^{(a,\alpha)}_{f,w,\ell,m}
        \;=\; \sum_{j\,\in\,\mathcal{V}_\alpha(a)} w(j)\; f(r_{ja})\;
        Y_\ell^{\,m}(\theta_{ja},\varphi_{ja})
\end{equation}
where $a$ stands for the central atom and $j$ for the embedding atoms
contained in an ensemble of neighbors up to $\alpha$,
$\mathcal{V}_\alpha(a)$, $Y_\ell^{\,m}$ are the real spherical harmonics,
$r_{ja}$ the distance between the atoms $a$ and $j$ and $f(r_{ja})$ the radial
function. Finally, $w(j)$ is an atomic-property descriptor of atom $j$.

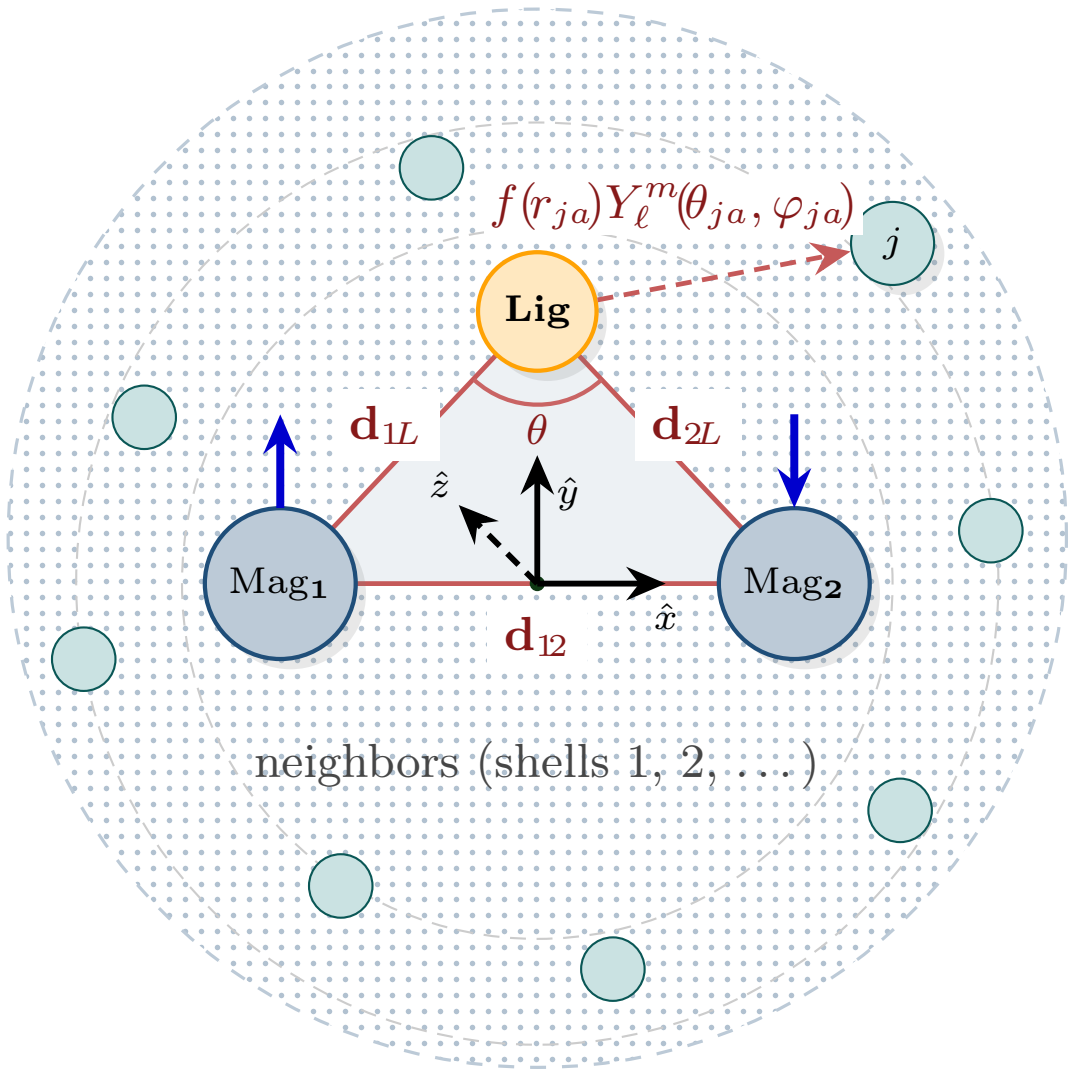
\begin{figure}[tbp]
\centering
\begin{tikzpicture}[
   >={Stealth[length=2.6mm,width=2.2mm]}, font=\small,
   mag/.style={circle, draw=verifyblue, line width=0.8pt, fill=verifyblue!30, minimum size=0.9cm, drop shadow={opacity=0.18}},
   lig/.style={circle, draw=cautionorange, line width=0.8pt, fill=cautionorange!25, minimum size=0.78cm, drop shadow={opacity=0.18}},
   env/.style={circle, draw=tealc!70!black, fill=tealc!22, minimum size=0.42cm},
   jat/.style={circle, draw=tealc!70!black, fill=tealc!22, minimum size=0.42cm, drop shadow={opacity=0.18}},
   ax/.style={->, line width=1.1pt, black}]
  \fill[pattern=dots, pattern color=verifyblue!35] (0,0.3) circle (3.5);
  \draw[verifyblue!30, dashed, line width=0.6pt] (0,0.3) circle (3.5);
  \draw[gray!40, densely dashed] (0,0) circle (2.35);
  \draw[gray!40, densely dashed] (0,0) circle (3.05);
  \node[text=gray!55!black, font=\small, align=center] (vlab) at (0,-1.2) {neighbors (shells 1, 2, \ldots)};
  \foreach \p in {{(-2.6,1.1)},{(-3.0,-0.5)},{(-1.3,-2.0)},{(0.5,-2.55)},{(2.4,-1.5)},{(3.0,0.35)},{(-0.7,2.75)}}
     \node[env] at \p {};
  \coordinate (M1) at (-1.7,0);
  \coordinate (M2) at (1.7,0);
  \coordinate (L)  at (0,1.8);
  \fill[verifyblue!8] (M1)--(M2)--(L)--cycle;
  \draw[alertred!75, line width=0.9pt] (M1)--(M2)--(L)--cycle;
  \node[font=\small, alertred!75!black, fill=white] at (0,-0.35)    {$\mathbf d_{1\!2}$};
  \node[font=\small, alertred!75!black, fill=white] at (-1.,1.05) {$\mathbf d_{1\!L}$};
  \node[font=\small, alertred!75!black, fill=white] at (1.,1.05)  {$\mathbf d_{2\!L}$};
  \draw[alertred!70, line width=0.8pt] (-0.42,1.35) arc (227:313:0.62);
  \node[alertred!75!black, font=\footnotesize] at (0,1.02) {$\theta$};
  \node[mag] at (M1) {\scriptsize Mag$_{\mathbf 1}$};
  \node[mag] at (M2) {\scriptsize  Mag$_{\mathbf 2}$};
  \node[lig] (Lg) at (L)  {\scriptsize \bf Lig};
  \draw[->, line width=1.5pt, spincol] (-1.7,0.5)  -- (-1.7,1.12);
  \draw[->, line width=1.5pt, spincol] ( 1.7,1.12) -- ( 1.7,0.5);
  \fill[okgreen!60!black] (0,0) circle (0.05);
  \draw[ax] (0,0) -- (0.85,0)     node[below, font=\scriptsize]{$\hat{x}$};
  \draw[ax] (0,0) -- (0,0.85)     node[right, font=\scriptsize, pos=0.7]{$\hat{y}$};
  \draw[ax, densely dashed] (0,0) -- (-0.52,0.52) node[above left=-3pt, font=\scriptsize]{$\hat{z}$};
  \node[jat] (j) at (2.35,2.25) {\scriptsize $j$};
  \draw[->, densely dashed, alertred!75, line width=0.9pt] (Lg) -- (j);
  \node[alertred!75!black, font=\small, fill=white, inner sep=1pt] at (0.9,2.5) {$f(\!r_{ja}\!)Y_\ell^m\!(\!\theta_{ja},\varphi_{ja}\!)$};
\end{tikzpicture}
\caption{Schematic representation of the descriptors engineering.  The atoms
  directly involved in the magnetic interactions are labeled Mag$_1$, Mag$_2$
  and Lig. They define an exchange plane and are described by internal
  coordinates, atomic numbers charges and spins. 
  Each neighboring atom $j$ of the  \emph{environment} ---~ dotted pattern~---
  is accounted for in the atomic density decomposition provided within the SOAP descriptors. }
\label{fig:descriptor}
\end{figure}

Fig.~\ref{fig:descriptor} provides a schematic representation of how the
descriptors are engineered to be directly related to the seek physics.

Among the tested hyper-parameters were the number of
neighbors  to the central atoms, $\alpha$, the choice of the radial function, $f$, the maximal $l$
and the atomic properties of the embedding atoms $w$.

\section{Results}
\label{sec:res}

The database was divided into three sets. The first set includes all data from
the \lumo\  compound. No training is done on this set that is kept for transferability
checks. The data on the \ybmo, \scmo, \dymo\ and \bcs\ compounds were randomly
divided into a training set including 70\% of the data, and a test set
including the remaining 30\%.

We tested nine variations of small DNN 3-5 hidden layers (HL), alternating
activation functions, one out of two always being a Sigmoid, starting with the
input layer. 
 Each of these models were tested in three versions with
64, 128 and 512 neurons and with two loss functions, namely the Mean
Square Error (MSE) and an home-made loss inspired from the internal energy in
statistical quantum mechanics (eq~\ref{eq:briceloss}). Such a function should
penalize the data with large errors at large inverse temperatures $\beta$
\begin{equation*}
  \Delta E_i = (J_i-\tilde J_i)^2 \quad ; \quad  Z=  \sum_{j=1}^Ne^{-\beta \,\Delta E_j}
\end{equation*}
\begin{equation}
  \label{eq:briceloss}
  {\cal L} = \dfrac{1}{N-1} \sum_{i=1}^N  \left(1- \frac{e^{-\beta\,  \Delta E_i}}{Z}\right)\,  \Delta E_i
\end{equation}
 With the different descriptor hyper-parameters it
resulted in about 35000 trainings.

In order to analyze these results and extract the best models, we used a
recursive K-means
clustering~\cite{Kmeanclustering_MacQueen1967,RecursiveKmeanclustering_Steinbach2000}
pipeline, with quality-based pruning, and the following performance metrics:
the best RMSE over all epochs for both the train and test sets, the RMSE over
the test set at the best train epoch and reversely, the harmonic mean of the
best RMSE on the train and test sets, and a composite score  built from the
weighted previously cited RMSE. In order to avoid artifacts in the clustering
process, we repeated it with slightly different settings, i.e. different
weights in the composite score and different rejection criteria. A group of
models came out of the clustering workflow independently of the settings,
insuring the stability of the choice.

\begin{figure}[tbp]
\centering
\resizebox{\linewidth}{!}{\includegraphics{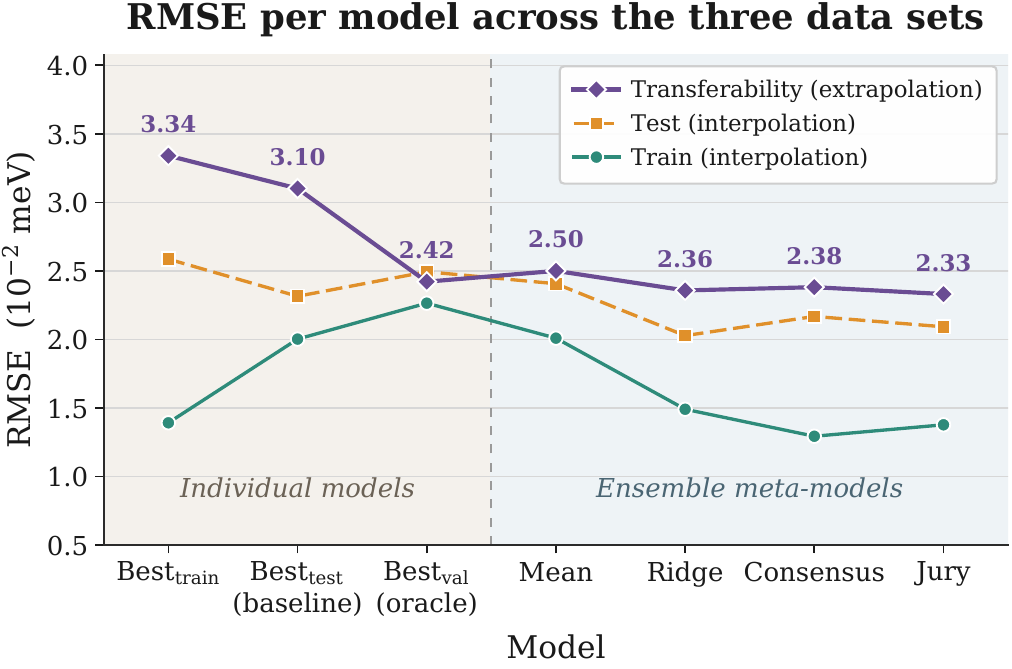}}
\caption{$10^{-2}\times$RMSE computed on the three datasets with the
  model performing best on the train set, the model performing best on the test
  set, the model performing best on the transferability set (never seen
  during the learning process) and the four ensemble models.}
\label{fig:rmse-compare}
\end{figure}

These selected models are then
aggregated using an ensemble meta-model in order improve the
performances. Different aggregation techniques were used (see
Fig.~\ref{fig:rmse-compare})~: the simple mean of the predicted $J$ values
over the selected models, the classical ridge
method~\cite{ens_ridge_Kennard_1970}, and two home-made methods referred to as
jury and  consensus.  In the consensus method, inspired from statistical physics, the
weights are chosen so that to minimize the error to the consensus.
\begin{eqnarray}
  \label{eq:ens}
  \check J_i &=&\sum_{m} \omega_m \, \tilde J_{i,m} \\ 
\text{and} \quad   \omega_m &=& \lambda_m e^{- \alpha \sum_i |J_i - \sum_{m'} \lambda_{m'} \tilde J_{i,m'}|} \\
\text{with}  \quad \sum_m \lambda_m  &=& 1
\end{eqnarray}
The $\tilde J_{i,m}$ and $\check J_i$ and then passed to a network with one hidden layer that learn 
$\alpha$, the $\lambda_m$ and the error to the consensus, $\Delta_i$,  using the $L_2$ loss function
\begin{eqnarray}
  \label{eq:cons_loss}
  {\cal L} &=& \sum_i \left(\check J_i+0.05\Delta_i - J_i \right)^2
\end{eqnarray}
$i$ refers to an observation, $J_i$ are the exact exchange integrals,
$\tilde J_{i,m}$ the predicted ones by the model $m$, $\check J_i$ the values
predicted by the ensemble meta-model. %
Finally, the Jury method, inspired from the Bagging
method\cite{Bagging_Breiman1996} and Bayesian committee~\cite{BC_Tresp2000},
consists in partitioning the selected models into 5 sets ${\cal M}_k$, on
which $\check J_i^{(k)}$ ensemble values are predicted. These
$\check J_i^{(k)}$ values, their median over the 5 sets and their standard
deviations over the sets are provided to a single neural network layer in
order to optimize the final $ \omega_m$.  Each ensemble meta-model is
leak-free trained, as trained on a 80\% subset of the original training set.

The results of the different ensemble meta-models are then evaluated on the
train and test sets and on the set kept for the transferability checks. The
latter only includes data on compounds never seen at any step of the
training. Fig.~\ref{fig:rmse-compare} presents the RMSE on the three data
sets, for the model best performing on the train set, on the test set and on
the transferability set, as well as the four ensemble meta-models. The first
important point is the global accuracy of all models. Indeed the RMSE is
around 0.7\% of the nominal $J$ values (\unit[$\simeq$3.5]{meV} in the three
sets). The second important point is that the models seem to correctly learn
the effect of the nature of the rare-earth counter-ion. Indeed,  even for the
best$_\text{train}$ and best$_\text{test}$ models the transferability accuracy
remain within 1\% error. Finally, the ensemble meta-models improve the results
over the three sets. The Consensus and Jury meta-models performing as well as
the best performing model on each data set simultaneously. The Jury meta-model
having slight advantage over the transferability check we will kept it as out
best model.

\begin{figure}[tbp]
\centering
\resizebox{8cm}{!}{\includegraphics{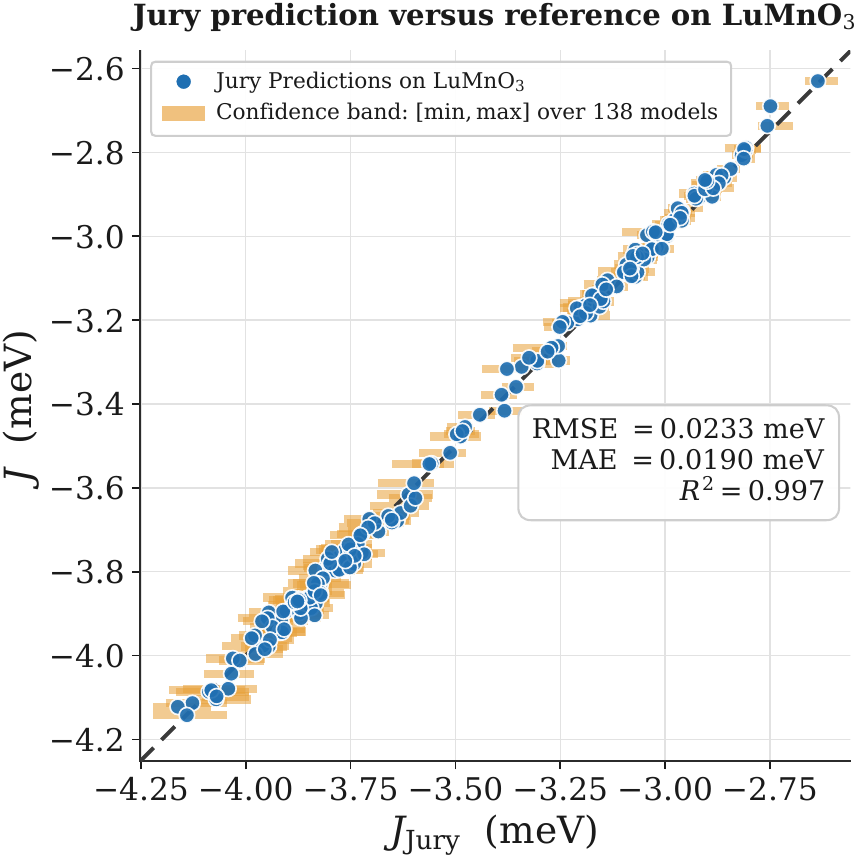}}
\caption{Performance of the Jury ensemble model on the transferability set (never seen during the training).}
\label{fig:rmse-jury}
\end{figure}

Figure~\ref{fig:rmse-jury} plots the Jury meta-model results on the
transferability data set. As one can see the model performs with a very good and
homogeneous accuracy over the whole $J$ range.

\subsection{Characteristics of the selected models}
\label{sec:hpsel}
Let us now analyze the 138 models selected by the clustering method. Eight
hyper-parameters were used, four of them related to the ML model, namely the
number of hidden layers, the size of the latent space, the activation and loss functions, and
four of them related to the embedding description (see Eq.~\ref{eq:SOAP}),
namely the the number of neighbors, $\alpha$ , the angular momentum maximum
value, $\ell_{\text{max}}$, the radial function, $f(r)$, and the weight function,
$w(j)$. Table~\ref{tab:hpsel} displays the representation ratio and the
enrichment factor of the different hyper-parameters in the selected models.

\begin{table}[htbp]
\caption{Representation ratio and enrichment factor for the different
  hyper-parameters tested and the 138 best models selected after clustering. 
 }
\begin{tabular}{lld{1}d{1}}
\hline\hline
  Hyper-parameter & Value & \multicolumn{1}{c}{Rep. ratio} & \multicolumn{1}{c}{Enrich.}  \\
  \hline 
\multirow{3}{*}{Hidden Layers}
                  & 3     & 52.2  & 1.59 \\
                  & 4     & 21.0  & 0.63 \\
                  & 5     & 26.8  & 0.79 \\[1ex]
\multirow{4}{*}{Latent space}
 & 512            & 34.1  & 1.29 \\
 & 256            & 29.0  & 1.00 \\
 & 128            & 27.5  & 0.95 \\
 & 64             & 9.4   & 0.61 \\[1ex]
\multirow{3}{*}{Activation}
 & GELU           & 62.3  & 1.96 \\
 & SiLU           & 33.3  & 0.95 \\
 & ReLU           & 4.3   & 0.13 \\[1ex]
\multirow{2}{*}{Loss}
 & MSE            & 100.0 & 1.25 \\
 & This work      & 0.0   & 0 \\[2ex]
\multirow{4}{*}{$\ell_{\text{max}}$}
 & 4        & 12.3  & 0.14 \\
 & 5        & 42.8  & 7.3  \\
 & 6        & 32.6  & 11.8 \\
 & 7        & 12.3  & 4.3  \\[1ex]
\multirow{2}{*}{$f(r)$}
 & Only $1/r^2$            & 74.6  & 1.50 \\
 & $\mathrm{const}+1/r^2$   & 25.4  & 0.50 \\[1ex]
  \multirow{3}{*}{\shortstack[l]{$w(j)$}}
 & Without ionic radius      & 0.0   & 0 \\
 & Ionic radius              & 17.4  & 0.94 \\
 & Ionic radius $+$ charge   & 82.6  & 2.16 \\[1ex]
\multirow{3}{*}{$\alpha$}
 & 1             & 21.7  & 1.29 \\
 & 1,2           & 21.0  & 1.23 \\
 & 1,2,3         & 19.6  & 1.17 \\
  \hline\hline
\end{tabular}
\label{tab:hpsel}
\end{table}

Let us first notice that the home-made physics inspired loss function does not
seem to work, while the standard MSE loss works well. More importantly the
ionic radius appears to be a crucial parameter as no model that does not
include it survives to the clustering. Going back to the chemistry, one shall
remember that the ionic radius of the rare-earth is what makes the difference
between the different compounds. Indeed, the energetic gap between the oxygen
and metal orbitals is strongly affected by the size of neighboring counter-ion, 
and as a result, the super-exchange component of $J$. The learning of this
effect is at the core of the present model transferability to unknown
compounds.

The second important point is that the best models are the narrowest (only 3
hidden layers), while the latent space size does not seem to matter much, provided it is
at least 128. Again this is understandable as we provided data as directly
related as possible to the physical effect we try to learn (\# HL) and the
embedding description involves a large number of data per observation (latent
space).  Finally alternating sigmoid and GELU activation functions seems to
perform best, probably due to its  continuity properties.

Coming to the embedding hyper-parameters one won't be surprised that the
maximal $\ell$ value should be large enough to account for the shape of the
embedding. In the present hexagonal manganites, the $c_{l,m}$ embedding data
of eq.~\ref{eq:SOAP} exhibit large values for $\ell=1,5,$ thus requiring
$\ell_{\text{max}} \ge 5$, but we expect that this is related to the specific
shape of the embedding and may be different for different structural
model. Finally as could be expected, the further away the embedding atoms are
the less they are important. A $1/r^2$ radial function insures to correctly
reproduce the local electric field around the atoms involved in the
interaction. Again this is physically informed as the local electric field is
of crucial importance in the screening of the raw super-exchange interactions.  

\section*{Conclusion}
\label{sec:concl}
Effective magnetic coupling is one of the key ingredient in the modeling of
low energy excitations in Mott's insulators. The accurate evaluation of such
integrals are demanding both in human and computational time thus perfect
example for the application of machine learning methods.

In this work we developed an ML method able to learn both the structure and
chemical nature of the counter-ions and predict magnetic exchanges within 1\%
accuracy. The key points of this method is to physics inform the ML model
through the appropriate data engineering. For this purpose we split the
crystal in a system containing the atoms involved in the seek magnetic
exchange, and an embedding. While the composition and structure of the system
part is provided through atomic number and internal coordinates, the embedding
information is provided through a SOAP decomposition~\cite{Bartok2013} with a
radial $\nicefrac{1}{r^2}$ penalty function.  Finally increased accuracy and
transferability is obtained through the use of an ensemble model after a
clustering to select the best models over the hyper-parameters and seeds. 

In further work we plan to extend the method to all transition-metal oxides
and chalcogenides Mott insulators as we believe our method contains the key
physical ingredients to describe this family of compounds.

\section*{Acknowledgments}
The authors thank the IDRIS and CINES computer center for providing computer
hours through the GENCI project N°0801842. THis work was funded by the ANR
``MagDesign'' (N°ANR-22-CE29-0018-01) and the MIAI cluster chair ``SpiNN''.

\bibliography{MagIA_biblio}
\end{document}